      [1999/12/01 v1.4c Il Nuovo Cimento]
\documentclass{cimento}

\title{Revisiting Caianiello's Maximal Acceleration {\mdseries\ttfamily}}
\author{G. ~Papini\from{ins:x} \from{ins:y}}
        \instlist{\inst{ins:x} Department of Physics, University of Regina, Regina, Sask.
        S4S 0A2, Canada
   \inst{ins:y} International Institute for Advanced Scientific Studies, 84019 Vietri sul Mare (SA),
  Italy}
\PACSes{\PACSit{03.65.-w, 03.65.Ta, 74.55.+h} {Quantum theory, maximal acceleration,
Ehrenfest theorem}}
\begin{document}

\maketitle

\begin{abstract}
A quantum mechanical limit on the speed of orthogonality evolution justifies the last
remaining assumption in Caianiello's derivation of the maximal acceleration. The limit is
perfectly compatible with the behaviour of superconductors of the first type.

\end{abstract}

\section{Introduction}
\label{1} Following previous attempts \cite{cai1}, Caianiello showed in 1984 that
Heisenberg's uncertainty relations place an upper limit  ${\mathcal A}_{m}$ on the value
that the acceleration of a particle can take along a worldline \cite{cai2}.  This limit,
referred to as maximal acceleration (MA), is determined by the particle's mass itself. It
is distinct from the value that has been derived in some works \cite{br,cai3,gasp} from
the Planck mass $ m_{P}= \left(\hbar c/G \right) ^{\frac{1}{2}}$ and is therefore a
universal constant. With some modifications \cite{wood} and additions, Caianiello's
argument goes as follows.

If two observables $ \hat{f}$ and $ \hat{g}$ obey the commutation relation
\begin{equation}\label{al}
\left[\hat{f},\hat{g}\right]= - i \hbar \hat{\alpha},
\end{equation}
where $ \hat{\alpha}$ is a Hermitian operator, then their uncertainties
\begin{eqnarray}
 \left(\Delta f\right)^{2}&=&
<\Phi\mid\left(\hat{f}-<\hat{f}>\right)^{2}\mid\Phi>\\ \nonumber
 \left(\Delta g\right)^{2}&=&
<\Phi\mid\left(\hat{g}-<\hat{g}>\right)^{2}\mid\Phi>
\end{eqnarray}
 also satisfy the inequality
\begin{equation}
\left(\Delta f\right)^{2}\cdot\left(\Delta g\right)^{2}\geq
\frac{\hbar^{2}}{4}<\Phi\mid\hat{\alpha}\mid\Phi>^{2},
\end{equation}
or
\begin{equation}\label{be}
\Delta f\cdot\Delta g\geq \frac{\hbar}{2} \mid<\Phi\mid\hat{\alpha}\mid\Phi>\mid .
\end{equation}
Using Dirac's analogy between the classical Poisson bracket $ \left\{f,g\right\}$ and the
quantum commutator \cite{land}
\begin{equation}
\left\{f,g\right\} \rightarrow\frac{1}{i\hbar}\left[\hat{f},\hat{g}\right],
\end{equation}
one can take $ \hat{\alpha}=\left\{f,g\right\}\hat{\mathbf{1}}$. With this substitution,
Eq.(\ref{al}) yields the usual momentum-position commutation relations. If in particular
$ \hat{f}=\hat{H}$, then Eq.(\ref{al}) becomes
\begin{equation}
\left[\hat{H},\hat{g}\right]= - i \hbar \left\{H,g\right\}\hat{\mathbf{1}},
\end{equation}
Eq.(\ref{be}) gives \cite{land}
\begin{equation}\label{&}
\Delta E\cdot \Delta g \geq \frac{\hbar}{2} \mid \left\{H,g\right\}\mid
\end{equation}
and
\begin{equation}\label{&&}
\Delta E\cdot\Delta g\geq \frac{\hbar}{2}\mid\frac{dg}{dt}\mid ,
\end{equation}
when $ \frac{\partial g}{\partial t}=0 $. Eqs.(\ref{&}) and (\ref{&&}) are re-statements
of Ehrenfest theorem. Criteria for its validity are discussed at length in the literature
\cite{mess,land,bal}. Eq.(\ref{&&}) implies that $\Delta E =0$ when the quantum state of
the system is an eigenstate of $\hat{H}$. In this case $\frac{dg}{dt}= 0$.

If $g \equiv v(t)$ is the (differentiable) velocity expectation value of a particle whose
average energy is $ E$, then Eq.(\ref{&&}) gives
\begin{equation}\label{7}
\mid \frac{dv}{dt}\mid \leq \frac{2}{\hbar} \Delta E \cdot \Delta v(t) .
\end{equation}
 In general \cite{sha}
\begin{equation}
 \Delta v =
\left(<v^{2}>-<v>^{2}\right)^{\frac{1}{2}}\leq v_{max}\leq c .
\end{equation}
 Caianiello's additional assumption, $ \Delta E \leq E$, has so far remained unjustified.
In fact, Heisenberg's uncertainty relation
\begin{equation}\label{yy}
\Delta E \cdot
\Delta t\geq \hbar/2 ,
\end{equation}
 that follows from (\ref{7}) by writing $ \Delta t =
\Delta v /|dv/dt| $, seems to imply that, given a fixed average energy E, a state can be
constructed with arbitrarily large $\Delta E$, contrary to Caianiello'assumption. This
conclusion is erroneous. The correct interpretation of (\ref{yy}) is that a quantum state
with spread in energy $ \Delta E$ takes a time $ \Delta t\geq \frac{\hbar}{2 \Delta E}$
to evolve to a distinguishable (orthogonal) state. This evolution time has a lower bound.
Margolus and Levitin have in fact shown \cite{marg} that the evolution time of a quantum
system with fixed average energy $E$ must satisfy the more stringent limit
\begin{equation} \label{&&&}
\Delta t\geq \frac{\hbar}{2E} ,
\end{equation}
which determines a {\it maximum speed of orthogonality evolution}  \cite{beck}. Their
argument is simple. If at $ t = 0$ an arbitrary state of a quantum system is written as a
superposition of energy eigenstates $ |\psi\left(0\right)> = \Sigma_{n}c_{n}|E_{n}> $,
then at time $ t$ the state has evolved to $ |\psi \left(t\right)> =
\Sigma_{n}c_{n}e^{-iE_{n}\pi t/\hbar}|E_{n}> $. The shortest time after which $ |\psi(0)>
$ and $ |\psi(t)> $ are distinguishable is given by the orthogonality condition
\begin{equation}\label{&&&&}
<\psi \left(0\right)|\psi \left(t\right)> =\Sigma_{n}|c_{n}|^{2} e^{-i\frac{E_{n}\pi
t}{\hbar}}= 0.
\end{equation}
The factor $  \pi $ in (\ref{&&&&}) has been introduced because (\ref{yy}) requires that
the energy distribution oscillate in time with a period at least $ \frac{\hbar}{2\Delta
E}$. On using the inequality $ cos(x)\geq 1-\frac{2}{\pi}(x + sin(x)) $, which is valid
for $ x \geq 0 $, and equating to zero both real and imaginary parts of (\ref{&&&&}),
Margolus and Levitin arrive at the equation
\begin{equation}\label{99}
Re\left(<\psi \left(0\right)|\psi
\left(t\right)>\right)=\Sigma_{n}|c_{n}|^{2}cos\left(\frac{E_{n}\pi\Delta
t}{\hbar}\right)\geq 1 - \frac{E \Delta t}{\hbar} ,
\end{equation}
 from which (\ref{&&&}) follows. Obviously, both
limits (\ref{yy}) and (\ref{&&&}) can be achieved only for $ \Delta E = E$, while spreads
$ \Delta E > E $, that would make $ \Delta t $ smaller, are precluded by (\ref{&&&}).
This effectively
 restricts $\Delta E$ to values
{\it$\Delta E \leq E$, as conjectured by Caianiello}. One can now derive an upper limit
on the value of the proper acceleration. In fact, in the instantaneous rest frame of the
particle, where the acceleration is largest \cite{wood}, $ E = mc^{2}$ and (\ref{7})
gives
\begin{equation}\label{8}
\mid \frac{dv}{dt}\mid \leq 2\frac{mc^{3}}{\hbar}\equiv {\mathcal A}_{m} .
\end{equation}

 It also follows that in the rest frame of
the particle, where $ \frac{d^{2}x^{0}}{ds^{2}}= 0 $, the absolute value of the proper
acceleration is \cite{wood,steph}
\begin{equation}\label{9}
\left(\mid
\frac{d^{2}x^{\mu}}{ds^{2}}\frac{d^{2}x_{\mu}}{ds^{2}}\mid\right)^{\frac{1}{2}}=
\left(\mid\frac{1}{c^{4}}\frac{d^{2}x^{i}}{dt^{2}}\mid \right)^{\frac{1}{2}}\leq
\frac{\mathcal{A}_{m}}{c^{2}} .
\end{equation}
Eq.(\ref{9}) is a Lorentz invariant. The validity of (\ref{9}) under Lorentz
transformations is therefore assured.

Result (\ref{&&&}) can also be used to extend (\ref{8}) to include the average length of
the acceleration $ <a>$. If, in fact, $ v(t)$ is differentiable, then fluctuations about
its mean are given by
\begin{equation}\label{del}
\Delta v \equiv v-<v>\simeq \left(\frac{dv}{dt}\right)_{0}\Delta
t+\left(\frac{d^{2}v}{dt^{2}}\right)_{0}\left(\Delta t\right)^{2}+...  .
\end{equation}
Eq.(\ref{del}) reduces to $ \Delta v\simeq \mid\frac{dv}{dt}\mid \Delta t =<a> \Delta t$
for sufficiently small values of $\Delta t$, or when $\mid\frac{dv}{dt}\mid$ remains
constant over $\Delta t$. Eq.(\ref{&&&}) then yields
\begin{equation}\label{ga}
<a> \leq \frac{2cE}{\hbar}
\end{equation}
and again (\ref{8}) follows.

Eq.(\ref{&&&}) is relevant to quantum geometry\cite{cai4,pap,scarp}, the entire subject
of maximal acceleration \cite{several} and the field of computation \cite{marg}. This
does not exhaust its usefulness. Its predictions and those of (\ref{7}) are compared, in
the example below, with the behaviour of a well known class of quantum systems.

\section{"Maximal Acceleration" in Type-I Superconductors}
\label{2} The static behavior of superconductors of the first kind is adequately
described by London's theory \cite{degennes}. The fields and currents involved are weak
and vary slowly in space. The equations of motion of the superelectrons are in this case
\cite{til}
\begin{eqnarray}\label{10}
\frac{D\vec{v}}{Dt}&=&\frac{e}{m}\left[\vec{E}+ \left(\frac{\vec{v}}{c}\times
\vec{B}\right)\right]\nonumber\\ &=& \frac{\partial \vec{v}}{\partial
t}+\left(\vec{v}\cdot \vec{\nabla}\vec{v}\right) .
\end{eqnarray}
On applying (\ref{&&}) to (\ref{10}), one finds
\begin{equation}\label{11}
\sqrt{\left(\frac{1}{2}\vec{\nabla}v^{2}-\vec{v}\times \left(\vec{\nabla}\times
\vec{v}\right)\right)^{2}}\leq \frac{2}{\hbar}\bigtriangleup E\cdot \bigtriangleup v  ,
\end{equation}
and again
\begin{eqnarray}\label{12}
\sqrt{\frac{1}{4}\left(\nabla_{i}v^{2}\right)^{2}+\frac{e}{m
c}\epsilon_{ijk}\left(\nabla^{i}v^{2}\right)v^{j}B^{k}+ \left(\frac{e}{m
c}\right)^{2}\left[v^{2}B^{2}-\left(v_{i}B^{i}\right)^{2}\right]} \leq\nonumber \\
\frac{2}{\hbar}\triangle E \bigtriangleup v  ,
\end{eqnarray}
where use has been made of London's equation
\begin{equation}\label{13}
\vec{\nabla}\times \vec{v}=-\frac{e}{m c}\vec{B} ,
\end{equation}
and $ \epsilon_{ijk}$ is the Levi-Civita tensor. Static conditions, $ \frac{\partial
v}{\partial t}=0 $, make (\ref{11}) and (\ref{12}) simpler. Eq.(\ref{10}) can be used to
express the acceleration in term of the quantities $ \vec{E}$ and $ \vec{B}$ that are of
more direct experimental and theoretical interest for this class of superconductors.
London's theory in the linear case predicts that $ \vec{E}=0$ in the superconductor.
Eqs.(\ref{11}) and (\ref{12}) can therefore be used to calculate an upper limit on $
\vec{E}$ in the nonlinear version of London's theory. It is also useful, for the sake of
numerical comparisons, to apply (\ref{12}) to the case of a sphere of radius $ R$ in an
external magnetic field of magnitude $ B_{0}$ parallel to the polar axis. This problem
has an obvious symmetry and can be solved exactly. The exact solutions of London's
equations for $ r\leq R $ are well-known \cite{london} and are reported here for
completeness. They are
\begin{equation}\label{14}
B_{r}=\frac{4 \pi}{\beta^{2}c}\frac{1}{r}\frac{1}{\sin
\theta}\frac{\partial}{\partial\theta}\left(\sin\theta j_{\varphi}\right)
\end{equation}
\begin{equation}\label{15}
B_{\theta}=-\frac{4 \pi}{\beta^{2}c}\frac{1}{r}\frac{\partial}{\partial r}\left(r
j_{\varphi}\right) ,
\end{equation}
\begin{equation}\label{16}
j_{\varphi}=nev_{\varphi}=\frac{A}{r^{2}}\left(\sinh \beta r - \beta r \cosh \beta
r\right)\sin\theta ,
\end{equation}
where $A=-\frac{c}{4\pi}\frac{3B_{0}}{2}\frac{R}{\sinh\beta r}$, $ n$ is the density of
superelectrons and $\beta=\left(\frac{4\pi n e^{2}}{m c^{2}}\right)^{\frac{1}{2}}$
represents the reciprocal of the penetration length. From (\ref{10}) and (\ref{11}) one
obtains
\begin{equation}\label{17}
\mid E_{r}\mid \leq \frac{\mid
v_{\varphi}B_{\theta}\mid}{c}+\sqrt{\left(\frac{2mc}{e\hbar}\right)^{2}\left(\Delta
E\right)^{2}\left(\Delta v\right)^{2}-\left(\frac{v_{\varphi}}{c}B_{r}\right)^{2}} .
\end{equation}
For a gas of fermions in thermal equilibrium $ \Delta E\sim \frac{3}{5}\mu$, $ \Delta
v\sim \frac{3}{2}\sqrt{\frac{\mu}{2m}}$ and the chemical potential behaves as $ \mu
\approx \epsilon_{F}-\frac{\left(\pi kT\right)^{2}}{12\epsilon_{F}}\approx
\epsilon_{F}\sim 4.5\times 10^{-12}erg $ for $ T $ close to the transition temperatures
of type-I superconductors. The reality of (\ref{17}) requires that $ \Delta E\geq
\mu_{B}B_{r}$, where $ \mu_{B}= \frac{e\hbar}{2mc}$ is the Bohr magneton, or that $
\frac{3}{5}\epsilon_{F}\geq \mu_{B}B_{r}$. This condition is certainly satisfied for
values of $ B_{r}\leq B_{c}$, where $ B_{c} $ is the critical value of the magnetic field
applied to the superconductor. From (\ref{17}) one also obtains
\begin{equation}\label{18}
\mid E_{r}\mid\leq
\frac{3}{2m}\left(\frac{\epsilon_{F}}{2}\right)^{\frac{1}{2}}\left[\frac{\mid
B_{\theta}\mid}{c}+\sqrt{\left(\frac{3\epsilon_{F}}{5\mu_{B}}\right)^{2}-\left(\frac{B_{r}}{c}
\right)^{2}}\right].
\end{equation}
More restrictive values for $ \Delta E$ and $ \Delta v$ can be obtained from $ B_{c} $ .
The highest value of the velocity of the superelectrons must, in fact, be compatible with
$ B_{c} $ itself, lest the superconductor revert to the normal state. This value is
approximately a factor $ 10^{3}$ smaller than that obtained by statistical analysis. The
upper value $ v_{0}$ of $ v_{\varphi}$ is at the surface. From $ \Delta E \leq
\frac{1}{2}m v_{0}^{2}, \Delta v \leq v_{0}$ and (\ref{17}) one finds that at the
equator, where $ B_{r}=0$, $ E_{r} $ satisfies the inequality
\begin{equation}\label{19}
\mid E_{r}\mid \leq \frac{v_{0}}{c}\left(\mid B_{\theta}\mid
+\frac{v_{0}^{2}}{2\mu_{B}}\right).
\end{equation}
For a sphere of radius $R=1 cm$ one finds $v_{0}\simeq 4.4\times 10^{4} cm/s$  and $
E_{r} \leq 69 N/C$. If no magnetic field is present, then (\ref{19}) gives $ E_{r}\leq
4.2 N/C $. On the other hand London's equation gives
\begin{equation}\label{**}
E_{r}=\frac{m}{2e}\frac{\partial v_{\varphi}^{2}}{\partial r}\simeq 0.32 N/C.
\end{equation}
The experimental work of Bok and Klein \cite{bok} agrees with (\ref{**}). The MA limits
(\ref{18}) and (\ref{19}) are therefore consistent with (\ref{**}) and its experimental
verification.

This research was supported by the Natural Sciences and Engineering Research Council of
Canada.

\end{document}